\documentclass{article}
\usepackage{jcappub}
\usepackage{graphicx}
\usepackage{color}
\usepackage{multirow}
\usepackage{amsmath}
\usepackage{amssymb}
\usepackage{natbib}
\usepackage{aas_macros}
\usepackage{verbatim}
\usepackage[bottom]{footmisc}

\usepackage[normalem]{ulem}

\title{Constraining the time evolution of dark energy, curvature and neutrino properties
with cosmic chronometers}

\author[1,2]{Michele Moresco,}
\author[3,4]{Raul Jimenez,}
\author[3,4,5]{Licia Verde,}
\author[1]{Andrea Cimatti,}
\author[2]{Lucia Pozzetti,}
\author[6]{Claudia Maraston,}
\author[6]{and Daniel Thomas} 

\affiliation[1]{ALMA Mater Studiorum - Universit\'a degli Studi di Bologna, Dipartimento di Astronomia, via Ranzani 1, I-40127, Bologna, Italy}
\affiliation[2]{INAF - Osservatorio Astronomico di Bologna, via Ranzani 1, 40127 Bologna, Italy}
\affiliation[3]{ICREA \& ICC-UB, University of Barcelona (IEEC-UB), Marti i Franques, 1, Barcelona 08028, Spain}
\affiliation[4]{Radcliffe Institute for Advanced Study, Harvard University, MA 02138, USA}
\affiliation[5]{Institute of Theoretical Astrophysics, University of Oslo, Oslo 0315, Norway}
\affiliation[6]{Institute of Cosmology and Gravitation, Dennis Sciama Building, University of Portsmouth, Burnaby Road, Portsmouth, PO1 3FX, UK}

\emailAdd{michele.moresco@unibo.it}
\emailAdd{raul.jimenez@icc.ub.edu}
\emailAdd{liciaverde@icc.ub.edu}
\emailAdd{a.cimatti@unibo.it}
\emailAdd{lucia.pozzetti@oabo.inaf.it}
\emailAdd{claudia.maraston@port.ac.uk}
\emailAdd{daniel.thomas@port.ac.uk}

\abstract{
We use the latest compilation of observational Hubble parameter measurements estimated with the differential evolution of cosmic chronometers, 
in the redshift range $0<z<2$, to place constraints on cosmological parameters. We consider the sample alone and in combination with other 
state-of-the art cosmological measurements: CMB data from the latest Planck 2015 release, the most recent estimate of the Hubble constant $H_{0}$, 
a compilation of recent Baryon acoustic oscillation data, and the latest type IA cosmological supernovae sample. Since cosmic 
chronometers are independent of the assumed cosmological model, we are able to provide constraints on the parameters that govern the expansion 
history of the Universe in a way that can be used to test cosmological models. In particular, we show that the latest measurements of $H(z)$ 
obtained with cosmic chronometer from the Baryon Oscillation Spectroscopic Survey (BOSS) survey provide enough constraining power in 
combination with Planck 2015 data to constrain the time evolution of dark 
energy, yielding constraints that are competitive with those obtained using Supernovae and/or baryon acoustic oscillations. From late-Universe
probes alone ($z<2$) we find that $w_0 = -0.9 \pm 0.18$ and $w_a = -0.5 \pm 1.7$, and when combining also Planck 2015 data we obtain $w_0=-0.98\pm 0.11$ 
and $w_a=-0.30\pm0.4$. In the theoretical framework, these new constraints imply that nearly all quintessence models are disfavoured by the data; only phantom models or a pure cosmological constant are favoured by the data. This is a remarkable finding as it imposes severe constraints on the nature of dark energy.
For the curvature our constraints are $\Omega_k = 0.003 \pm 0.003$, considering also CMB data. 
We also find that $H(z)$ data from cosmic chronometers are important to constrain parameters that do no affect directly the expansion history, 
by breaking or reducing degeneracies with other parameters. We find that 
$N_{\rm eff} = 3.17 \pm 0.15$, thus excluding the possibility of an extra (sterile) neutrino at more than $5 \sigma$, and put competitive limits on the sum 
of neutrino masses, $\Sigma m_{\nu}< 0.27$ eV at 95\% confidence level. Finally, we constrain the redshift evolution of dark energy by exploring separately 
the early and late-Universe, and find a dark energy evolution $w(z)$ consistent with the $\Lambda$CDM model at the 40\% level over the entire redshift range $0 < z < 2$.}

\begin{document}

\maketitle


\section{Introduction}

Currently, the $\Lambda$CDM model represents the simplest framework
to describe all available cosmological information. Within this model, the Universe has no spatial
curvature, the present day energy is mostly constituted of a dark energy component in the form of
a cosmological constant, and there are three massless neutrinos. This model can accurately match
at the $\sim$percent level current observations with the minimal number of parameters.

One approach to make progress in understanding the nature of dark energy is trying to measure quantities 
that are independent of the cosmological model. One such technique is to measure directly the expansion 
history of the Universe: this can be done using massive and passively evolving early-type galaxies as ``cosmic chronometers'' \cite{Jimenez2002}, 
thus providing {\it standard(-izable) clocks} in the Universe.
The basic idea underlying this approach is based on the measurement
of the differential age evolution as a function of redshift of these chronometers, which provides a direct estimate
of the Hubble parameter $H(z)=-1/(1+z)dz/dt \simeq -1/(1+z)\Delta z/\Delta t$.
The main strength of this approach is the reliance on the measurement of a differential quantity, 
$\Delta z/\Delta t$, which provides many advantages in minimizing many common issues 
and systematic effects (for a detailed discussion, see \cite{Moresco2012a}). There has been significant progress both in the theoretical 
understanding of the method and control of systematics uncertainties \cite{Moresco2011,Moresco2012b} and in the improvement of 
observational data \cite{Moresco2012a,Moresco2015,Moresco2016}. 

Because the cosmic chronometer method measures directly the expansion history, the most interesting parameters to be constrained 
are those affecting the background evolution, chiefly the evolution of dark energy and the curvature. However, other parameters which 
from Cosmic Microwave Background data alone show degeneracies with the background evolution, such as neutrino properties, 
are also affected.

The main aim of this paper is to explore what constraints, independently of the cosmological model (i.e. $\Lambda$CDM-inferred parameters), 
can be obtained from the the new dataset of $H(z)$ measurements from Ref.~\cite{Moresco2016} both alone and in combination with other data. 
In particular, we focus on constraints on the time evolution of dark energy, and we demonstrate how early and late-Universe probes give a 
consistent picture. Further, we demonstrate that the new $H(z)$ data, in combination with the other late-Universe probes, allow for a reconstruction up to 3 free parameters of the time evolution of dark energy, when dark energy is parameterised via a Chebyshev expansion. 
We also explore the standard Chevallier-Polarski-Linder (CPL) parameterization, providing accurate constraints on dark energy evolution.
Despite so much freedom in the form of dark energy as a function of time, we find it to be consistent with a cosmological constant, with only deviations of 
about 40\% permitted in the redshift range $0 < z <2$. 
The rest of the paper is organised as follows. In \S \ref{sec:data} we present the data, both the $H(z)$ data and the other state-of-the-art cosmological 
data and in \S \ref{sec:theory} we state our underlying assumptions and the methodological approach. In \S \ref{sec:results} we present our results, 
first for the cosmic chronometers data alone, then in comparison with external data sets and finally in combination, discussing our constraints on dark
energy time evolution, curvature, and neutrino properties.We conclude in \S \ref{sec:conclusions}.


\section{Data}
\label{sec:data}

In this analysis, we compare the constraints on cosmological parameters that can be obtained by probes that map the late-time Universe ($z<2$) expansion history,
as well as the improvement that can be obtained by combining those with early-time Universe probes.
The baseline of our analysis is the Hubble parameter measurements obtained with the cosmic chronometers (hereafter CC) technique, but we
consider as well more ``standard'' probes such as Supernovae Type Ia (hereafter SNe), Baryon Acoustic Oscillation distance measurements 
(hereafter BAO), and local $H_0$ measurements (hereafter $H_0$).
Early-time Universe probes are the latest Cosmic Microwave Background (CMB) measurements from the Planck mission (hereafter Planck15). 
We use directly the posterior sampling provided by the Planck collaboration for specific models and for the full combination of temperature and 
polarisation power spectrum data (in the Planck15 nomenclature  TT,TE,EE+lowP)\footnote{Downloadable from \texttt{http://pla.esac.esa.int/pla/}\#\texttt{cosmology} 
under ``Full grid of results"}. We refer to Ref. \cite{Planck2015} for more information. 
In the following we present the other datasets considered.

\begin{figure}[t!]
\begin{center}
\begin{minipage}{0.98\textwidth}
\centering
\includegraphics[width=8.25cm]{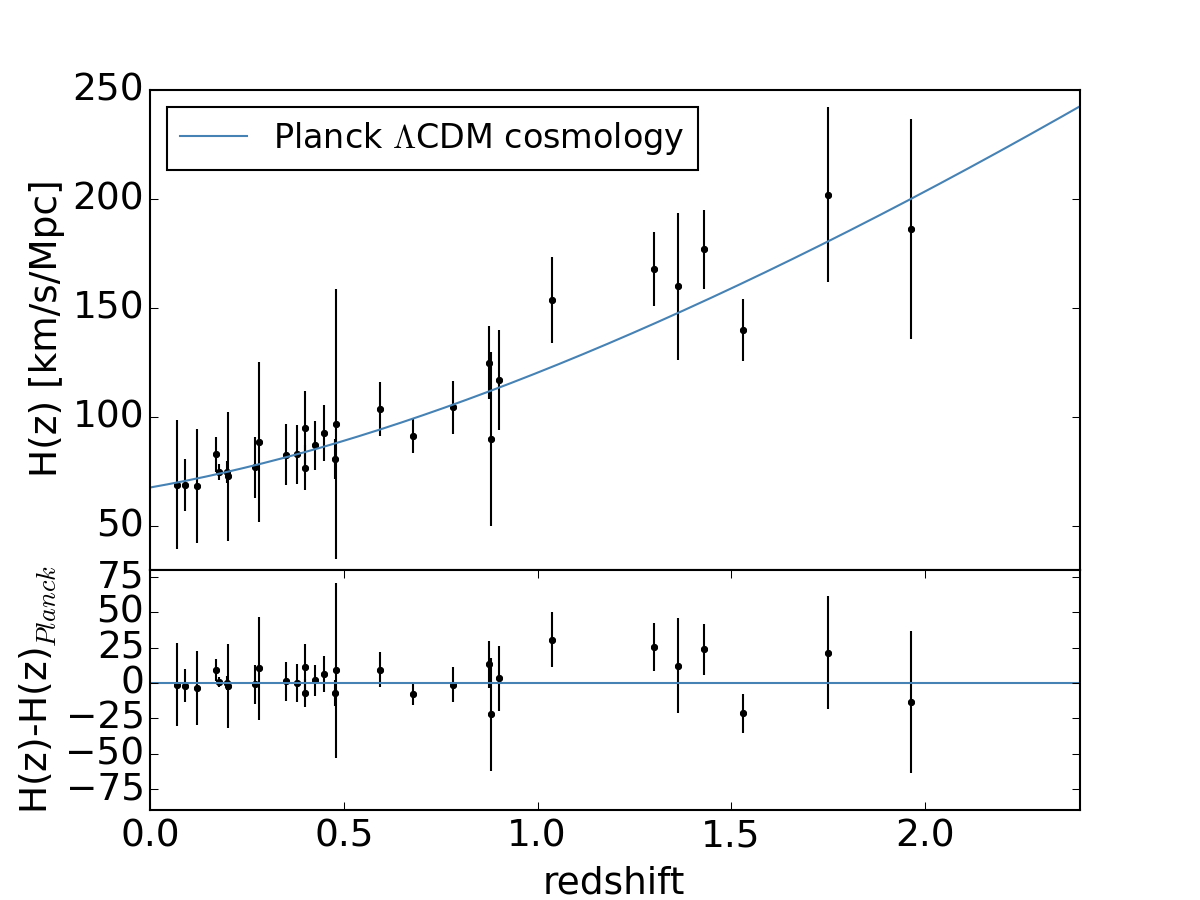}
\end{minipage}
\begin{minipage}{0.49\textwidth}
\centering
\includegraphics[width=8.25cm]{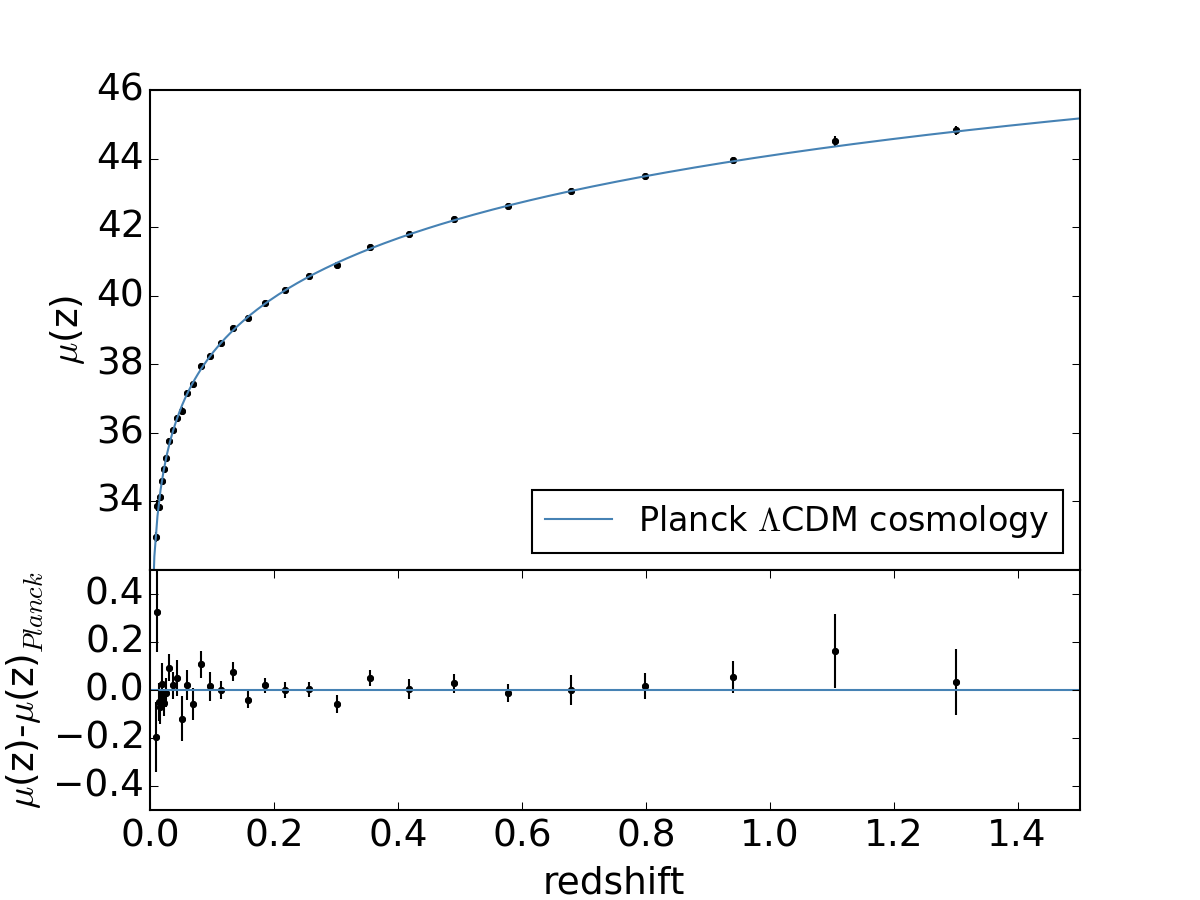}
\end{minipage}
\begin{minipage}{0.49\textwidth}
\centering
\includegraphics[width=8.25cm]{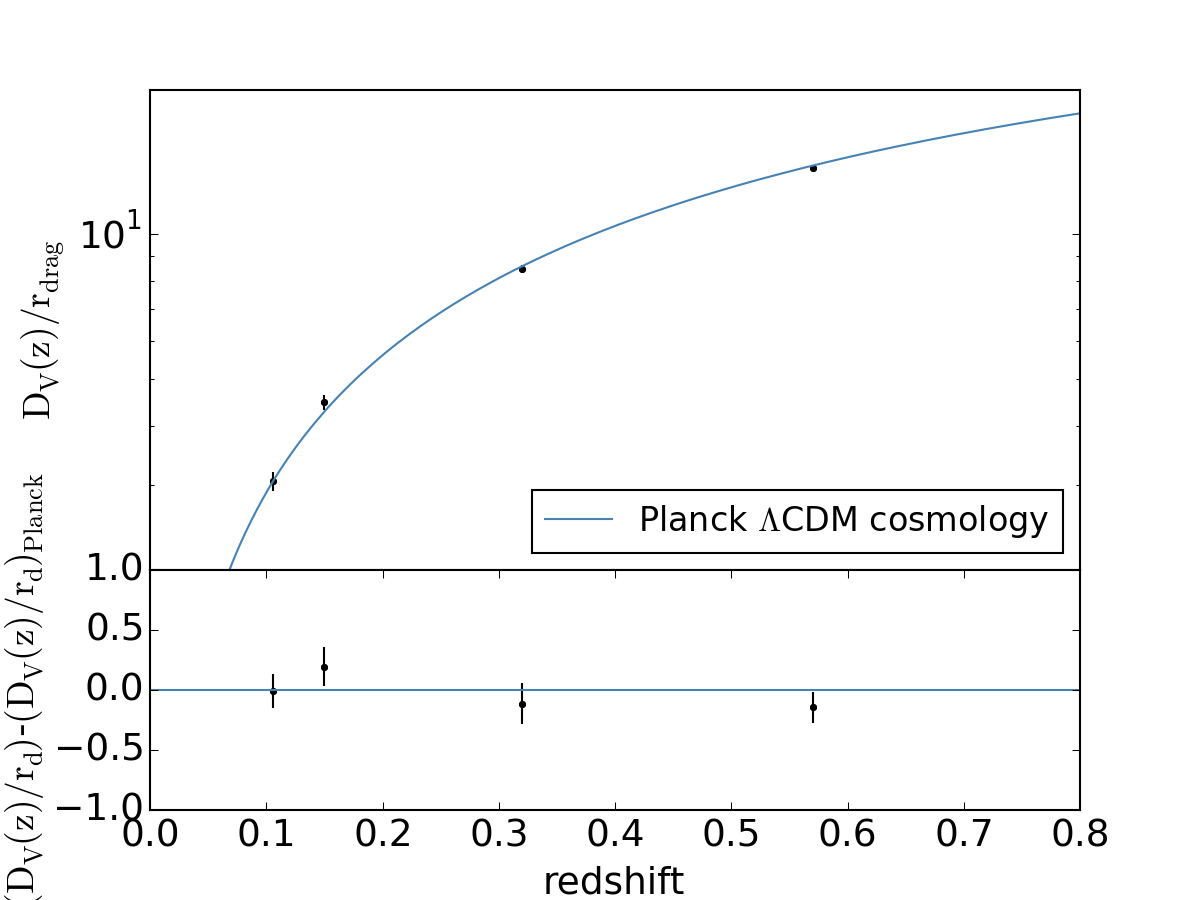}
\end{minipage}
\caption{Late-time expansion history dataset used in this analysis. The blue lines do not represent a fit to the data, but show the fiducial Planck 
$\Lambda$CDM cosmology ($H_{0}=67.8$ km/s/Mpc, $\Omega_{m}=0.308$). Lower panels show the residuals
of the data with respect to the fiducial Planck cosmology. CC data have been taken from Refs. 
\cite{Simon2005,Stern2010,Moresco2012a,Zhang2014,Moresco2015,Moresco2016}, SNe data from Ref.~ \cite{Betoule2014}, and
BAO data from Refs.~\cite{Beutler2011,Ross2015,Anderson2014}.
\label{fig:data}}
\end{center}
\end{figure}

\subsection{Cosmic chronometer dataset}

The {\it cosmic chronometers} approach to measure $H(z)$ was first introduced in Ref.~\cite{Jimenez2002}; it uses relative ages of 
the most massive and passively evolving galaxies to measure $dz/dt$, from which $H(z)$ is inferred. The latest implementation has been 
explained in detail in Ref.~\cite{Moresco2012a}, where the possible sources of uncertainty and related issues are also discussed; we refer 
to those references for a comprehensive discussion.
We consider the compilation of Hubble parameter measurements provided by \cite{Moresco2016}. It contains the latest updated list of $H(z)$ 
measurements \cite{Simon2005,Stern2010,Moresco2012a,Zhang2014,Moresco2015,Moresco2016} obtained with the cosmic chronometers approach, comprising 
of 30 measurements spanning the redshift range $0<z<2$. This sample covers roughly 10 Gyr of cosmic time; the data are presented in Fig. \ref{fig:data}.
The CC approach to measure $H(z)$ has the desirable feature of being largely independent on assumptions about the cosmological model (besides isotropy and homogeneity); it does, however, rely on the identification of an optimal tracer of the aging of the Universe with redshift (a cosmic chronometer), 
and the reliable dating of its age. An extended discussion can be found in Ref.~\cite{Moresco2016}. Of particular importance is a possible dependence on the 
adopted evolutionary stellar population synthesis (EPS) model, which is key in determining the age of the chronometer. In this analysis
we used the measurements calibrated on \cite{BC03} (BC03) EPS models, since they provide the largest dataset to date. However, 
Ref.~\cite{Moresco2012a,Moresco2015,Moresco2016} provided measurements also with the newest M11 \cite{MaStro} models for a smaller dataset. 
We explore the dependence of our results on the adopted EPS model in Appendix \ref{sec:modeldep}.

\subsection{Additional datasets}
More standard cosmological probes have been also exploited as complementary datasets to this analysis.

{\bf Type 1A supernovae (standard candles)} We consider the latest ``joint light curves'' (JLA) sample \cite{Betoule2014}, comprising 740 SNe Ia from the three year
Sloan Digital Sky Survey, Supernova Legacy Survey \cite{Astier2006,Sullivan2011}, Hubble Space Telescope 
\cite{Riess2007,Suzuki2012} and other local experiments (see \cite{Conley2011}). Here we use the binned
distance modulus provided by Ref.~\cite{Betoule2014}, with its associated covariance matrix. It is defined as:
\begin{equation}
\mu_{b}=M+5\log_{10}D_{L}(z)
\end{equation}
where M is a ({\it nuisance}) normalization parameter and $D_{L}(z)$ the luminosity distance at redshift $z$. The luminosity distance at redshift $z$ is related to an 
integral of the Hubble parameter from redshift $0$ to $z$. As such, it offers sensitivity to the curvature parameter but its integral nature makes it less sensitive 
than CC to sharp variations in $H(z)$. Also marginalisation over $M$ makes this probe sensitive to the shape of $H(z)$ but not to its overall normalisation 
(characterised for example by the value at a given redshift like $z=0$, $H_0$).

{\bf Baryon Acoustic oscillation (standard rulers)}
Our BAO analysis is based on the (isotropic) acoustic-scale distance ratio $\rm D_{V}(z)/r_{drag}$ where $r_{\rm drag}$ is the sound horizon at radiation drag,
\begin{equation}
D_{V}(z)=\left[(1+z)^{2}D_{A}^{2}(z)\frac{cz}{H(z)}\right]^{1/3}\,,
\label{eq:DVtheor}
\end{equation}
and $D_{A}(z)$ is the angular diameter distance at redshift $z$.

Our compilation comprises the measurements obtained by 6dFGS \cite{Beutler2011}, SDSS Main Galaxy Sample \cite{Ross2015} and BOSS LOWZ and CMASS surveys
\cite{Anderson2014} and is similar to the baseline BAO dataset used in Planck 2015 analysis \cite{Planck2015}.
$D_V$ is a combination of an integral of $H(z)$ (through $D_A$) and a direct $H(z)$ measurement \footnote{Anisotropic BAO measurements can 
measure $D_A$ and $H(z)$ separately  but at present the anisotropic measurements does not yet have significantly more statistical 
power than the isotropic one \cite{Anderson2014}. While  Ref.\cite{Planck2015} for CMASS uses the anisotropic measurement from \cite{Anderson2014}, here we use the isotropic one. }, but the exquisite measurement of $r_{\rm drag}$ provided by CMB data offers a tight constraint on the overall 
normalisation of the relation (see e.g., discussion in \cite{Cuesta:2014asa, Aubourg:2014yra}). 
BAO measurements have been provided also for other surveys (e.g. Wigglez \cite{Kazin2014}), or for other subsamples (e.g. galaxy clusters, e.g 
\cite{Veropalumbo2014,Veropalumbo2016}), but since the covariance between these samples and the dataset used in this analysis has not been
estimated, we decided not to use them.

{\bf Local $H_0$ value.} Finally in some cases we also include the local value of $H_0$ as measured by Ref.~\cite{Riess2011} using the recalibration of 
Ref.~\cite{Humphreys2013}, which yields $H_0=73.0\pm 2.4$ km/s/Mpc. Since there have been claims of tensions between the local $H_0$ measurement and 
the value inferred from early-Universe observations within a $\Lambda$CDM Universe (e.g., \cite{Efstathiou:2013via, Verde:2013wza, planck, Planck2015} 
and references therein) when including CMB data we will always report results both with and without the $H_0$ measurement. 
In this way the reader can judge wether the inclusion of a possible ``discrepant" measurement drives any of the conclusions.

The SNe and BAO data used in this analysis are shown in Fig.\ref{fig:data}.


\section{Methodology}
\label{sec:theory}

Under the assumptions of isotropy and homogeneity, the metric of space-time of the Universe is fully specified by the Friedmann-$\mathrm{Lema\hat{\i}tre}$-
Robertson-Walker one. The most economic $\Lambda$CDM cosmological model can be described by six parameters, but three of them univocally define the 
background evolution described by the Hubble parameter $H(z)=H_{0}\sqrt{\Omega_{\rm m}(1+z)^{3}+\Omega_{\rm DE}}$.
Extensions to this model have been proposed by relaxing one or more of its assumptions, such as flatness, dark energy Equation-of-State (EoS) parameter 
evolution, number of relativistic species in the Universe, total sum of neutrino masses; in this context, a generic model for the expansion rate of the Universe 
adopts a generic form for the equation of state parameter of dark energy $w(z)$, and is:
\begin{equation}
H(z)=H_{0}\left\{\Omega_{r}(1+z)^4+\Omega_{m}(1+z)^{3}+\Omega_{k}(1+z)^{2}+\Omega_{\rm DE}(1+z)^{3(\int_0^z\frac{w(z')}{(1+z')}dz')}\right\}^{1/2}\,.
\label{eq:Hztheor0}
\end{equation}
where $\Omega_{i}$ denote the energy density parameter for the various species in the Universe (matter, curvature, dark energy 
and radiation) at $z=0$.
Given that the contribution to the total energy due to radiation is not significant at late time where $\Omega_{\rm DE}$ is important, we can safely neglect it; 
the relation between the energy density parameters is $1=\Omega_{m}+\Omega_{\rm DE}+\Omega_{k}$.

A popular model for the expansion rate of the Universe is given by the CPL \cite{Chevallier2001,Linder2003} parameterisation of the equation of state for 
dark energy, $w(z)=w_{0}+w_{a}(z/(1+z))$, yielding for the expansion rate: 
\begin{equation}
H(z)=H_{0}\left\{\Omega_{r}(1+z)^4+\Omega_{m}(1+z)^{3}+\Omega_{k}(1+z)^{2}+\Omega_{\rm DE}(1+z)^{3(1+w_{0}+w_{a})}e^{-3w_{a}\frac{z}{1+z}}\right\}^{1/2}\,.
\label{eq:Hztheor}
\end{equation}
 
The stronger dependence of $H(z)$ is on the Hubble constant, matter density and curvature parameter, while the dependence on dark
energy EoS parameters is less significant, particularly for $w_{a}$ which is the most difficult parameter to be constrained. 

We follow two approaches. First, we analyze the constraints that can be put on cosmological parameters, and in particular on dark energy time 
evolution, with late-Universe probes (i.e. at $z<2$, not considering CMB data), such as CC, SNe and BAO (with and without $H_0$), first separately 
then combined; then we explore how the constraints 
can be narrowed down by adding early-Universe information from CMB observations. Therefore, we will be able to test how the late and early Universe agrees 
with each other, being this fully determined within a given cosmological model, like $\Lambda$CDM.
 
We analyze the goodness of fit of CC, SNe and BAO with a standard $\chi^{2}$ approach. This is possible as the reported errors on the data are Gaussianly distributed. 
We sample the distribution of parameters with the public python package \texttt{emcee} 
\cite{emcee}, which is an implementation of the affine-invariant ensemble sampler for Markov chain Monte Carlo (MCMC) proposed by Ref.~\cite{Goodman2010}. 
For CC we have that:
\begin{equation}
\chi^{2}=\sum\frac{(H_{\rm th}(z)-H_{\rm obs}(z))^{2}}{\sigma_{H_{\rm obs}(z)}^{2}},
\end{equation}
where $H_{\rm th}(z)$ is taken from Eq. \ref{eq:Hztheor} and $H_{\rm obs}$ and $\sigma_{H_{\rm obs}}$ from Ref.~\cite{Moresco2016}. 

For SNe we have:
\begin{equation}
\chi^{2}=(\mu_{b}-M-5\log_{10}(D_{L}(z))^{t}C^{-1}(\mu_{b}-M-5\log_{10}(D_{L}(z)),
\end{equation}
where $D_{L}$ is the luminosity distance and $C$ the covariance matrix associated with distance modulus measurements $\mu_{b}$
\footnote{http://supernovae.in2p3.fr/sdss\_snls\_jla/ReadMe.html}.

For BAO we have:
\begin{equation}
\chi^{2}=\sum\frac{(D_{\rm V,th}(z)/r_{\rm drag}-D_{\rm r,obs}(z))^{2}}{\sigma_{D_{\rm r,obs}(z)}^{2}},
\end{equation}
where $D_{r}=D_{V}/r_{\rm drag}$,  and $D_{\rm V,th}$ is given by Eq. \ref{eq:DVtheor}. When CMB information is not included,
$r_{\rm drag}$ is treated as a nuisance parameter  i.e., marginalised over.
 
We considered uniform priors on the following variables: $H_{0}=[50,100]$ km/s/Mpc, $\Omega_{m}=[0.01,0.99]$, $w_{0}=[-3,0]$, $w_{a}=[-5,5]$,  $r_{\rm drag}=[100-200]$.

We first explore a flat model ($\Omega_{k}=0$) with the CPL parameterisation for $w(z)$ (fw$_{0}w_{a}$CDM hereafter) to compare the results with the ones that can 
be obtained from CMB \cite{Planck2015}.

Subsequently, to include early time (CMB) constraints we have used the posterior samples provided by the 
Planck 2015 data release, and importance sampled them with CC measurements, comparing the results with the ones obtained from the
combination of Planck15+BAO and Planck15+SNe provided by the Planck team \cite{Planck2015}.

\section{Results}
\label{sec:results}

\begin{figure}[t!]
\begin{center}
\includegraphics[width=0.49\textwidth]{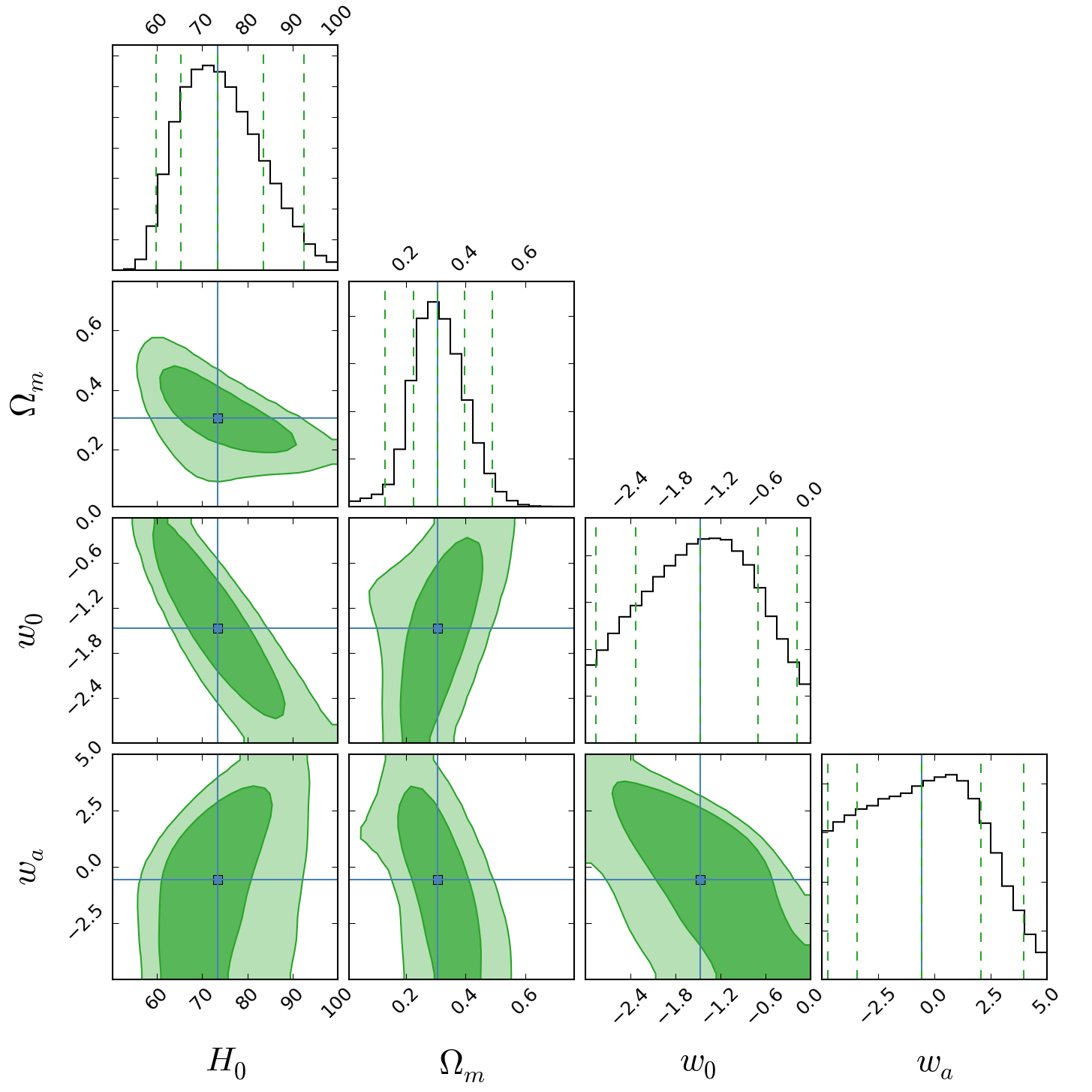}
\includegraphics[width=0.49\textwidth]{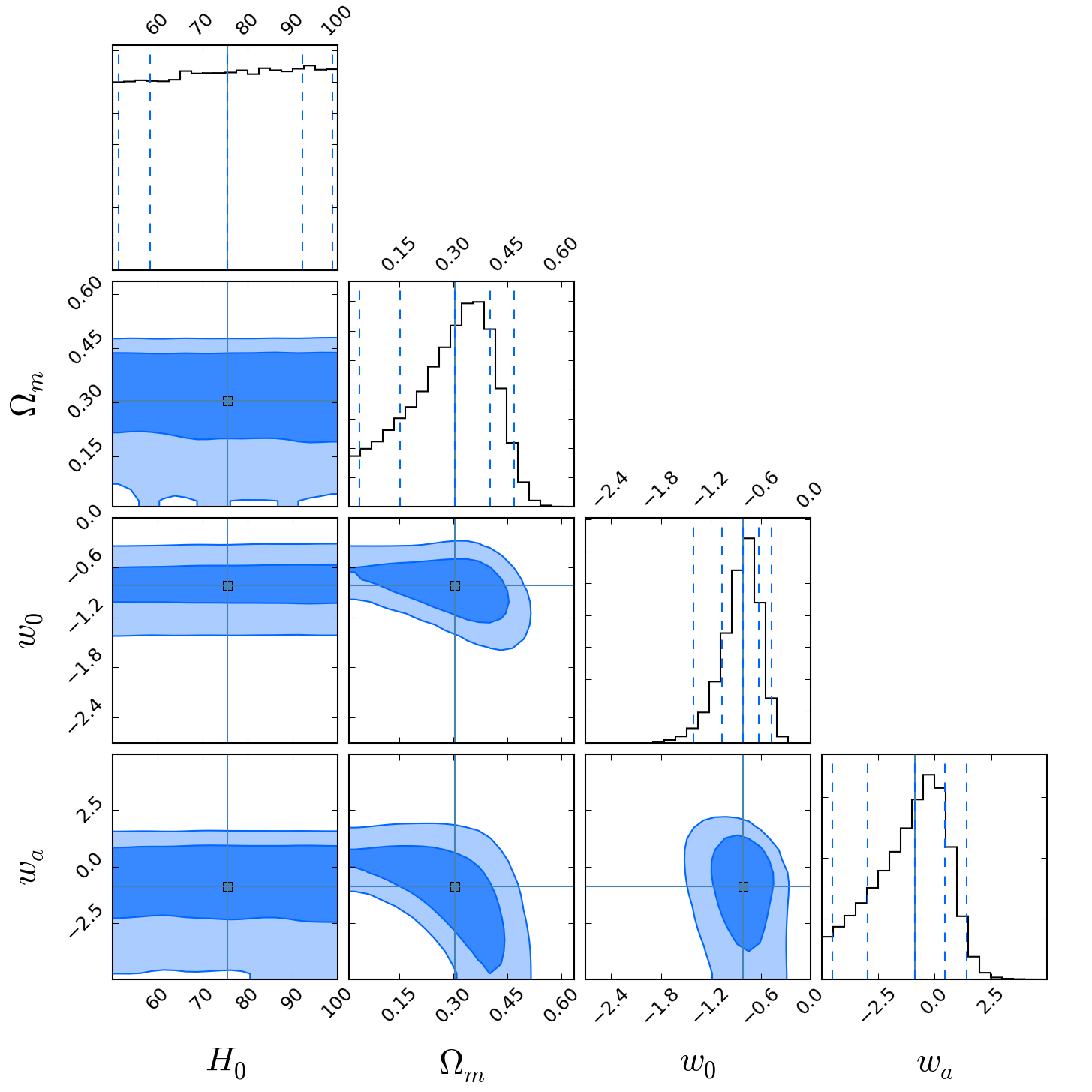}
\includegraphics[width=0.49\textwidth]{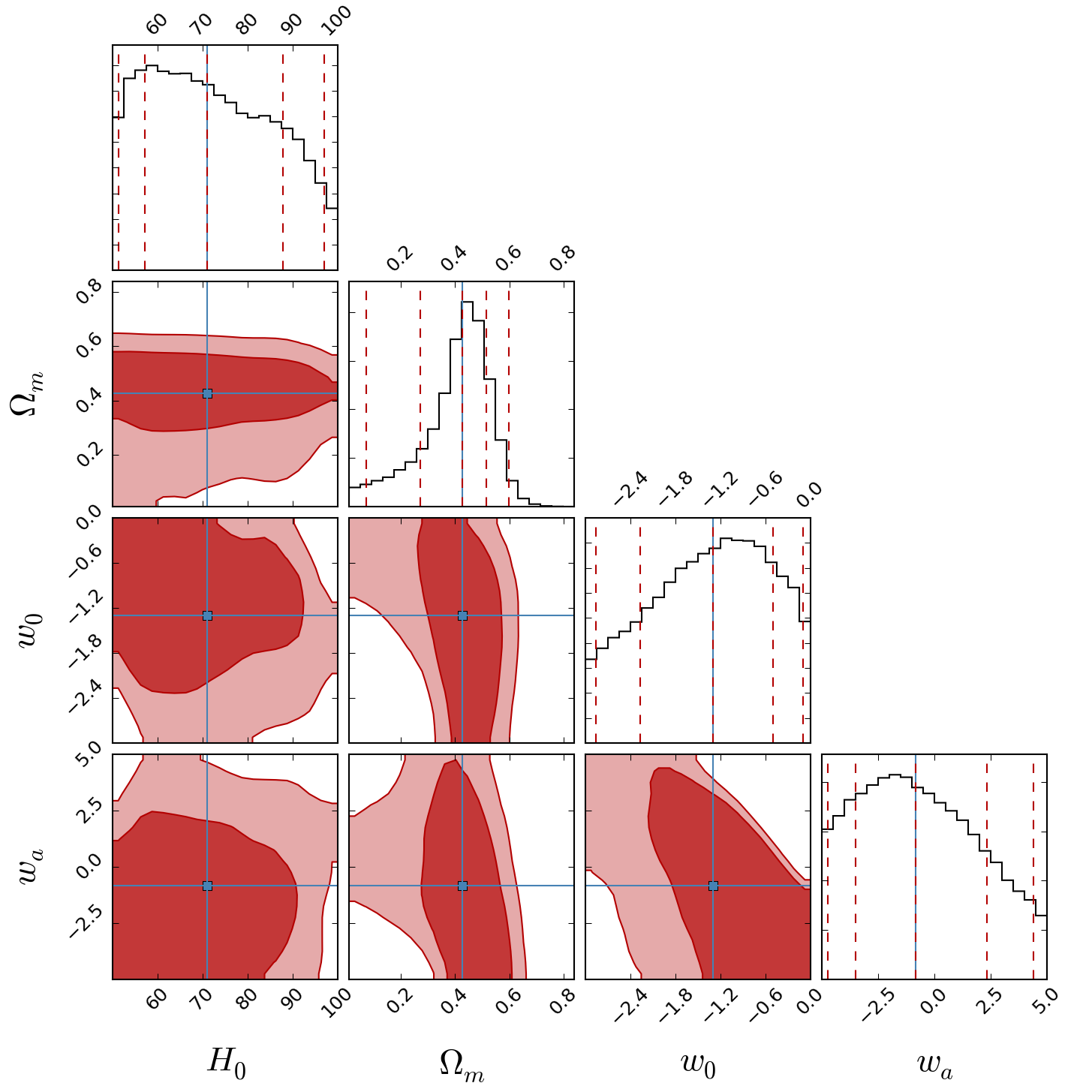}
\includegraphics[width=0.49\textwidth]{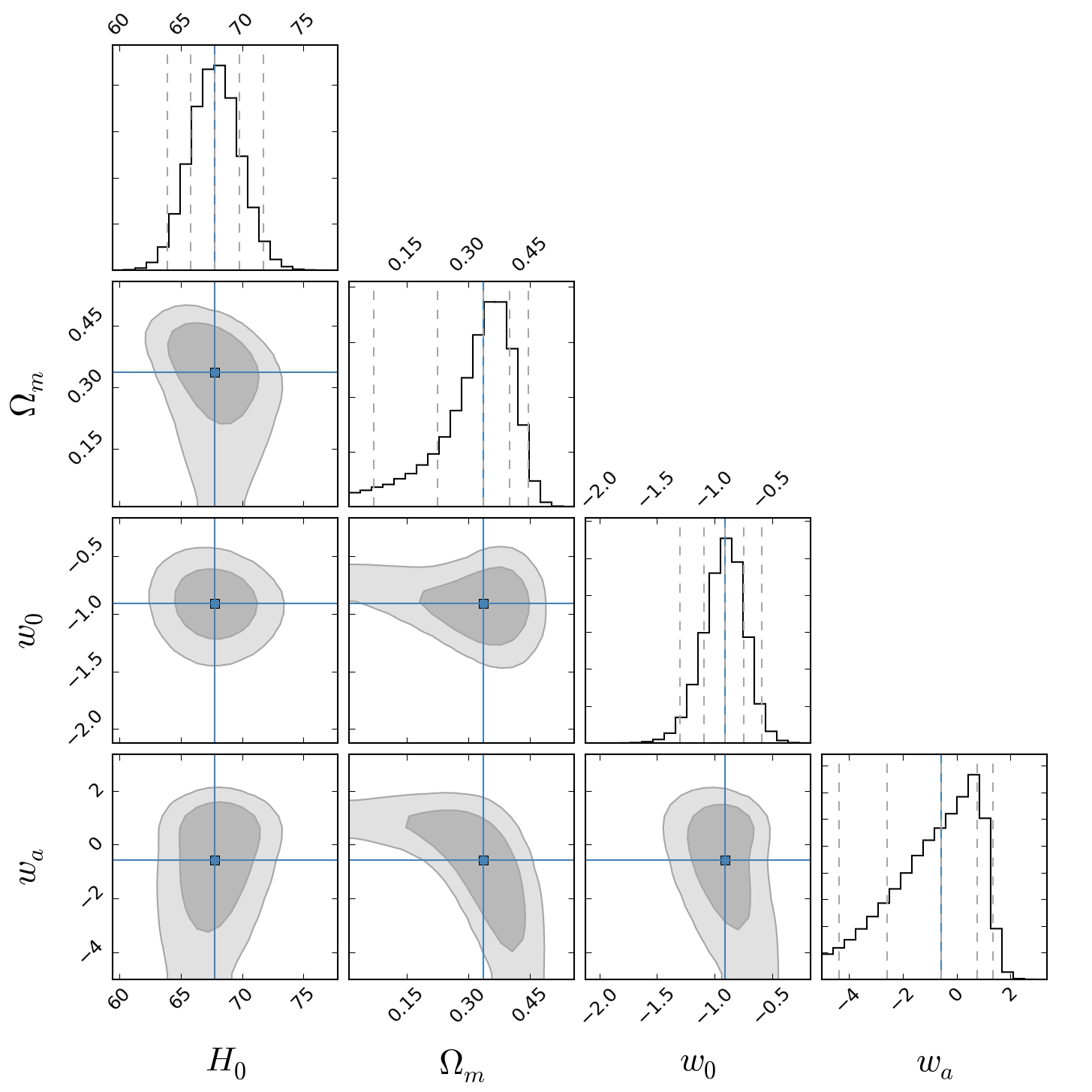}
\caption{Constraints on a flat $w_{0}w_{a}$CDM cosmology from cosmic chronometers (in green, upper-left panel), SNe (in blue, 
upper-right panel), BAO data (in red, lower-left panel), and from the combination of the various probes (in grey, lower-right panel) 
obtained with a MCMC approach. In each panel the contour plots are shown at 68\% and 95\% confidence level, and the posterior distribution 
of H$_{0}$, $\Omega_{m}$, $w_{0}$ and $w_{a}$, with the 68\% and 95\% confidence level limits.
\label{fig:fw0waCDM}}
\end{center}
\end{figure}

We start by discussing the results obtained from late-Universe probes: cosmic chronometer analysis alone, and then in combination with other probes (BAO, SNe). 
We then combine late-Universe probes with early-Universe observations (CMB) to improve constraints also on parameters that do not have a direct effect on $H(z)$, 
such as the number of the relativistic species in the Universe $N_{\rm eff}$ and the sum of the neutrino masses $\Sigma m_{\nu}$, by breaking parameter degeneracies.

\subsection{Constraints from the late-Universe probes}
\label{sec:reslate}

\begin{table}[b!]
\begin{center}
\begin{tabular}{ccccc}
\multicolumn{5}{c}{\small MARGINALIZED 1D CONSTRAINTS}\\
\multicolumn{5}{c}{f$w_{0}w_{a}$CDM model}\\
\hline \hline
& H$_{0}$ & $\Omega_{\rm m}$ & $w_{0}$ & $w_{a}$\\
\hline
CC & $73.2^{+10}_{-8.1}$ & $0.31^{+0.09}_{-0.08}$ & $-1.46^{+0.77}_{-0.87}$ & $-0.5^{+2.6}_{-2.9}$\\
SNe & -- & $0.31^{+0.1}_{-0.15}$ & $-0.82^{+0.19}_{-0.26}$ & $-0.9^{+1.3}_{-2.1}$\\
BAO & $70^{+17}_{-14}$ & $0.42^{+0.09}_{-0.16}$ & $-1.3^{+0.81}_{-0.97}$ & $-0.8^{+3.2}_{-2.7}$\\
\hline
CC+SNe+BAO & $67.8\pm2$ & $0.33^{+0.06}_{-0.12}$ & $-0.9^{+0.17}_{-0.19}$ & $-0.5^{+1.3}_{-2}$\\
\hline
\end{tabular}
\end{center}
\caption{Constraints on H$_{0}$, $\Omega_{\rm m}$, $w_{0}$ and $w_{a}$ (68\% confidence limits) obtained for a flat $\Lambda$CDM 
cosmology with equation-of-state parameter for dark energy parameterised as $w(z)=w_{0}+w_{a}(z/(1+z))$.}
\label{tab:fw0waCDM}
\end{table}

Fig. \ref{fig:fw0waCDM} shows constraints on a flat $w_{0}w_{a}$CDM cosmology using CC, SNe, BAO individually and their combination.
Marginalised constraints for each parameter are reported in Tab \ref{tab:fw0waCDM}.
The poor constraining power on $H_0$ of the BAO analysis results from the fact that $r_{\rm drag}$ is treated as a nuisance parameter 
(i.e. is left as a free parameter, without any CMB prior, and is marginalised over).
From this analysis, it is evident that CC and BAO have a similar constraining power in the $w_{0}-w_{a}$ plane, with CC providing 
better constraints on $H_{0}$ and slightly better constraints on $\Omega_{m}$; SNe, while insensitive to $H_{0}$, provide stringent constraints on the dark energy
EoS evolution. Most interestingly, the directions of the degeneracy between parameters in the various probes are different, and hence their
combination can provide more stringent constraints, as shown by Fig. \ref{fig:fw0waCDM} bottom right panel. 
More quantitatively, the constraints obtained with CC are $\sim$70\% and $\sim50$\% better than the ones from BAO respectively for $H_{0}$ and $\Omega_{m}$, 
while just $\sim$8\% for the dark energy EoS parameters. The constraints from SNe are better than the ones for CC of a factor $\sim3.5$ for $w_{0}$,
and of $\sim$60\% for $w_{a}$, while being worst by $\sim$ 50\% on $\Omega_{m}$.
The combination of all late-Universe probes yields to a significant improvement of $\sim40$\% on $\Omega_{m}$with respect to SNe alone, and helps to
partially remove the degeneracy in the $w_{0}-w_{a}$ plane, improving by $\sim$ 25\% the measurements of $w_{0}$.

In order to gain further insights into the time evolution of dark energy, it is useful to explore beyond the two-parameter $w_0-w_a$ fit. 
Ideally, one would like to reconstruct $w(z)$ in a non-parametric way; unfortunately, this is not possible (e.g. see Ref.~\cite{Simon2005}). 
So, alternatively, one needs to use a parameterisation that is the least possible model dependent. Ref.~\cite{Simon2005} proposed an expansion 
with Chebyshev polynomials, $T_i$, without any further assumptions on the shape of the dark energy EoS evolution:
\begin{equation}
w(z)=\sum_{i=0}^{N}\omega_{i}T_{i}(x(z))
\label{eq:Cheb1}
\end{equation}
where $T_{i}(x)$ are the Chebyshev polynomials: up to the second order $T_{0}=1$, $T_{1}=x$, $T_{2}=2x^{2}-1$,
$x=(2 z/z_{max})-1$ and $\omega_i$ are the parameters to be constrained by the data.
\begin{figure}[t!]
\begin{center}
\includegraphics[width=0.9\textwidth]{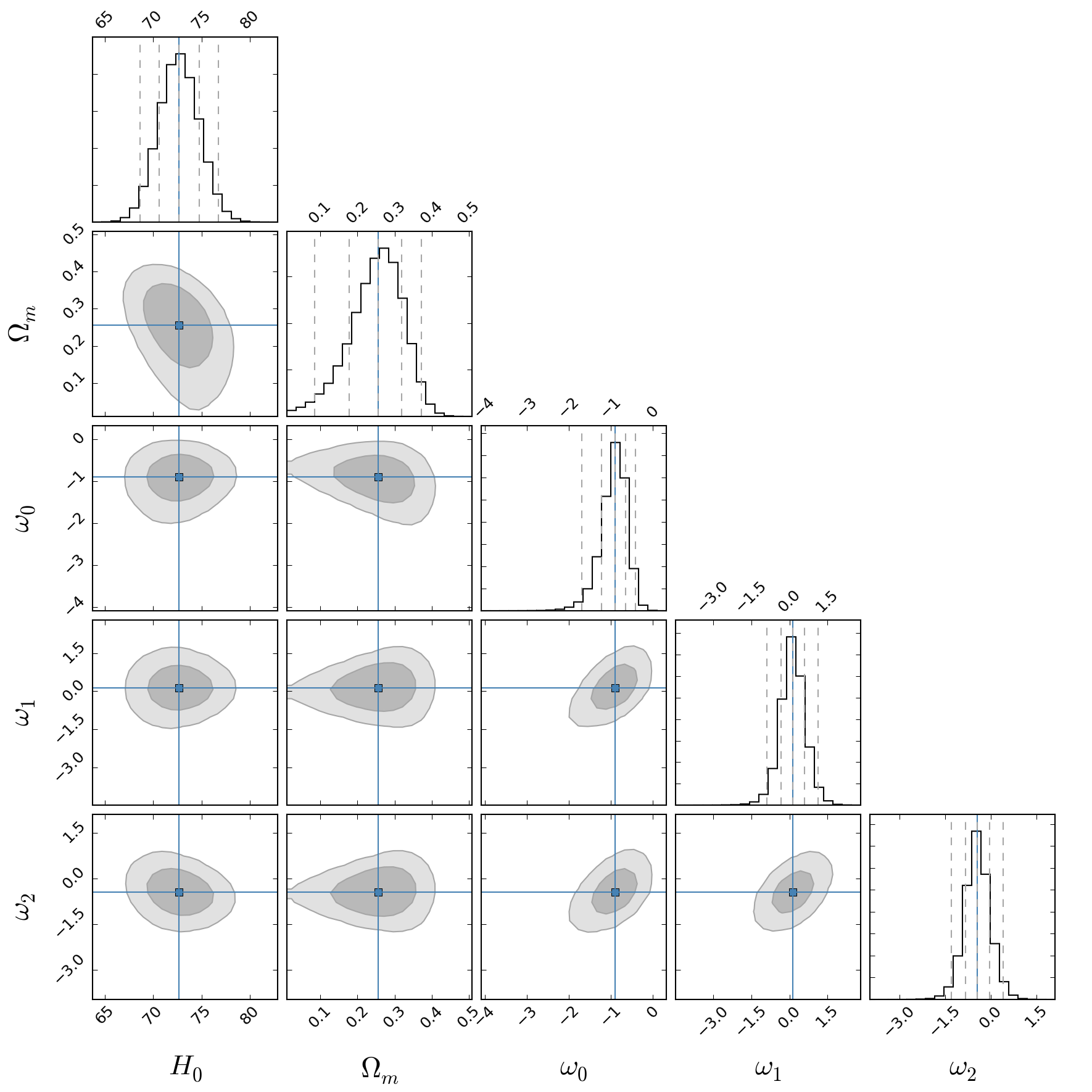}
\caption{Constraints for the Chebyshev expansion fit obtained from the combination of cosmic chronometer, SNe and BAO data. 
In each panel the contour plots are shown at 68\% and 95\% confidence level, and the posterior distribution 
of H$_{0}$, $\Omega_{m}$, $\omega_{0}$, $\omega_{1}$ and $\omega_{2}$, with the 68\% and 95\% confidence level limits.
For these constraints we assumed a Gaussian prior on $H0=73\pm2.4$ 
\cite{Riess2011,Humphreys2013,Cuesta2015}.
\label{fig:Cheb}}
\end{center}
\end{figure}

\begin{table}[t!]
\begin{center}
\begin{tabular}{ccccc}
\multicolumn{5}{c}{\small MARGINALIZED 1D CONSTRAINTS}\\
\multicolumn{5}{c}{Chebyshev expansion}\\
\hline \hline
& $\Omega_{\rm m}$ & $\omega_{0}$ & $\omega_{1}$ & $\omega_{2}$\\
\hline
CC+H$_{0}$ & $0.36\pm0.04$ & $-7.6^{+6.2}_{-6.5}$ & $-6.1^{+8.7} _{-9.3}$ & $-0.2^{+6.3}_{-3.5}$ \\
\hline
CC+SNe+BAO+H$_{0}$ - 1$\rm^{st}$ order  & $0.27^{+0.05}_{-0.06}$ & $-0.68^{+0.16}_{-0.21}$ & $0.53^{+0.30}_{-0.21}$ & -- \\
CC+SNe+BAO+H$_{0}$ - 2$\rm^{nd}$ order & $0.26^{+0.06}_{-0.08}$ & $-0.91^{+0.26}_{-0.33}$ & $0.13^{+0.47}_{-0.45}$ & $-0.46^{+0.4}_{-0.39}$ \\
\hline
\end{tabular}
\end{center}
\caption{Constraints on H$_{0}$, $\Omega_{\rm m}$, and $\omega_{0}$, $\omega_{1}$ and $\omega_{2}$ parameters (from Eq.~\ref{eq:Cheb1}) at  68\% confidence level 
obtained from the fit to the Chebyshev expansion of $w(z)$ up to the second order as modeled in Eq.~\ref{eq:Cheb2}. For these constraints
we assumed a Gaussian prior on $H_0=73\pm2.4$ \cite{Riess2011,Humphreys2013,Cuesta2015}, and hence do not provide the constraints for $H_{0}$.}
\label{tab:Cheb}
\end{table}
In this framework, it is possible to write the expansion history (for a flat Universe) as:
\begin{equation}
H(z)=H_{0}(1+z)^{3/2}\sqrt{\Omega_{m}+(1-\Omega_{m})\exp \left ( \frac{3}{2}z_{max}\sum_{i=0}^{N}\omega_{i}G_{i}(z) \right )}
\label{eq:Cheb2}
\end{equation}
where the functions $G_{i}$ can be obtained iteratively from the equation $J_{i}=1/nb[(2z/z_{max}-1)^{i}-(-1)^{i}]-(a/b)J_{i-1}$,
$J_{0}=1/b\log(1+z)$, $a=1+(z_{max}/2)$ and $b=z_{max}/2$. Up to the first two orders, it can be obtained that $G_{0}=J_{0}$,
$G_{1}=J_{1}$ and $G_{2}=2J_{2}-J_{0}$ \cite{Simon2005}.

In Tab.~\ref{tab:Cheb} we report the results on the constraints of $\omega_{i}$ up to both first and second order obtained by fitting jointly 
CC, SNe and BAO data to Eq.~\ref{eq:Cheb2}, and in Fig.~\ref{fig:Cheb} we show the contours for the fit up to the second order. 
Because of the increased freedom in the EoS parameterisation, in this analysis we always include the $H_0$ measurement.

The present-day value of $w$ in this expansion can be estimated as:
\begin{equation}
w_{z=0}=\sum_{i=0}^{N}(-1)^{i}\omega_{i}
\end{equation}
from which we find from the combined constraints $w_{z=0}=-1.21^{+0.69}_{-0.81}$ with the first order, 
and $w_{z=0}=-1.5^{+1.4}_{-1.56}$ with the second order expansion.

In this analysis by combining SNe and BAO data with the latest CC measurements, we are able to close for the first time the 2$\sigma$ 
contours on the $\omega_{i}$ parameters up to the second order, significantly improving on previous measurements \cite{Simon2005}
where unbounded constraints were provided, and only up to the first order. This is an important result, since it provides a measurement of the
evolution of dark energy EoS with a completely independent parameterisation. In Fig. \ref{fig:wz} we compare the estimated evolution
of $w(z)$ with different parameterisations and different datasets combinations, finding, remarkably, results in very good agreement, 
and compatible with a cosmological constant.


\subsection{Combining early-Universe information}
\label{sec:researly}

\begin{figure}[t!]
\begin{center}
\includegraphics[width=0.9\textwidth]{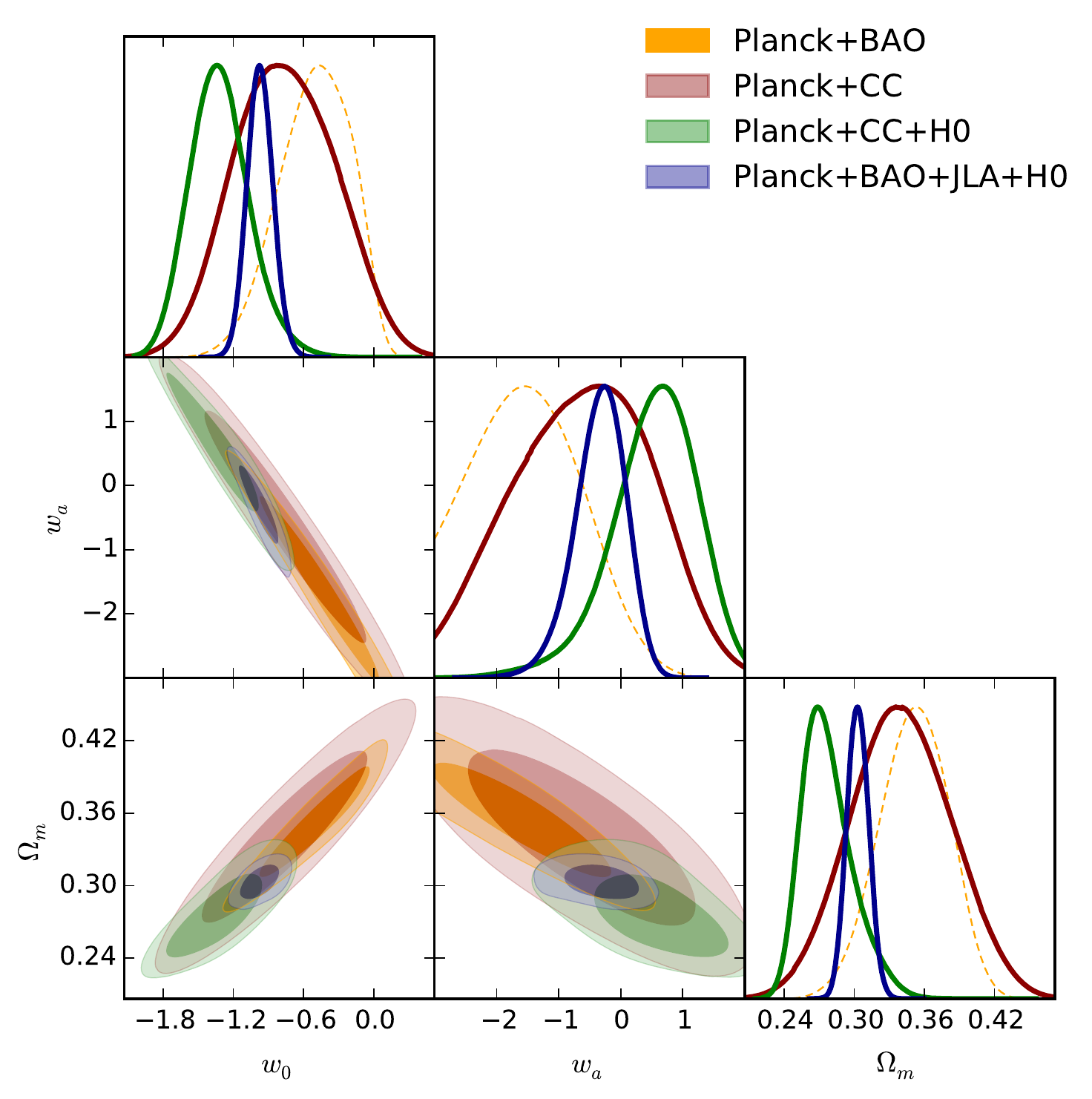}
\caption{Constraints on $\Omega_{\rm m}$, $w_0$ and $w_a$ for a flat $w_{0}w_{a}$CDM model 
obtained with different combinations of data sets. Cosmic chronometers and BAO have a similar
constraining power in the $w_{0}$-$w_{a}$ plane, as shown in Fig. \ref{fig:fw0waCDM},
and the combination of CMB+BAO and CMB+CC provides comparable results.
We used $H_{0}73\pm2.4$ km/s/Mpc from Ref.~\cite{Riess2011,Humphreys2013,Cuesta2015}.
\label{fig:fig1}}
\end{center}
\end{figure}

\begin{table}[b!]
\begin{center}
\begin{tabular}{cccc}
\multicolumn{4}{c}{\small MARGINALIZED 1D CONSTRAINTS}\\
\multicolumn{4}{c}{f$w_{0}w_{a}$CDM model}\\
\hline \hline
& $\Omega_{\rm m}$ & $w_{0}$ & $w_{a}$\\
\hline
Planck15+BAO & $0.351^{+0.032}_{-0.028}$ & $-0.51^{+0.36}_{-0.26}$ & $-1.47^{+0.78}_{-1.0}$ \\
Planck15+CC & $0.314^{+0.039}_{-0.052}$ & $-0.81\pm 0.40$ & $-0.8^{+1.2}_{-1.0}$ \\
Planck15+CC+$H_{0}$ & $0.273^{+0.017}_{-0.020}$ & $-1.18^{+0.20}_{-0.25}$ & $0.01^{+0.63}_{-0.44}$ \\
Planck15+BAO+CC+$H_{0}$ & $0.294^{+0.016}_{-0.018}$ & $-1.09^{+0.19}_{-0.21}$ & $-0.02^{+0.61}_{-0.49}$ \\
Planck15+BAO+SNe+$H_{0}$ & $0.3031\pm 0.0093$ & $-0.97\pm 0.11$ & $-0.34^{+0.45}_{-0.35}$ \\
Planck15+BAO+CC+SNe+$H_{0}$ & $0.3032\pm 0.0092$ & $-0.98\pm 0.11$ & $-0.30^{+0.42}_{-0.34}$ \\
\hline \hline
\end{tabular}
\end{center}
\caption{Marginalised constraints on $\Omega_{\rm m}$, $\Omega_{\rm DE}$, $\Omega_{\rm k}$, $w_{0}$ and $w_{a}$ at  68\% confidence level 
obtained for a open $\Lambda$CDM 
cosmology with equation-of-state parameter for dark energy parametrized as $w(z)=w_{0}+w_{a}(z/(1+z))$. We adopted $H_{0}=73\pm2.4$ 
km/s/Mpc from Ref.~\cite{Riess2011,Humphreys2013,Cuesta2015}.}
\label{tab:fw0waCDM_Pl}
\end{table}

\begin{figure}[t!]
\begin{center}
\includegraphics[width=\textwidth]{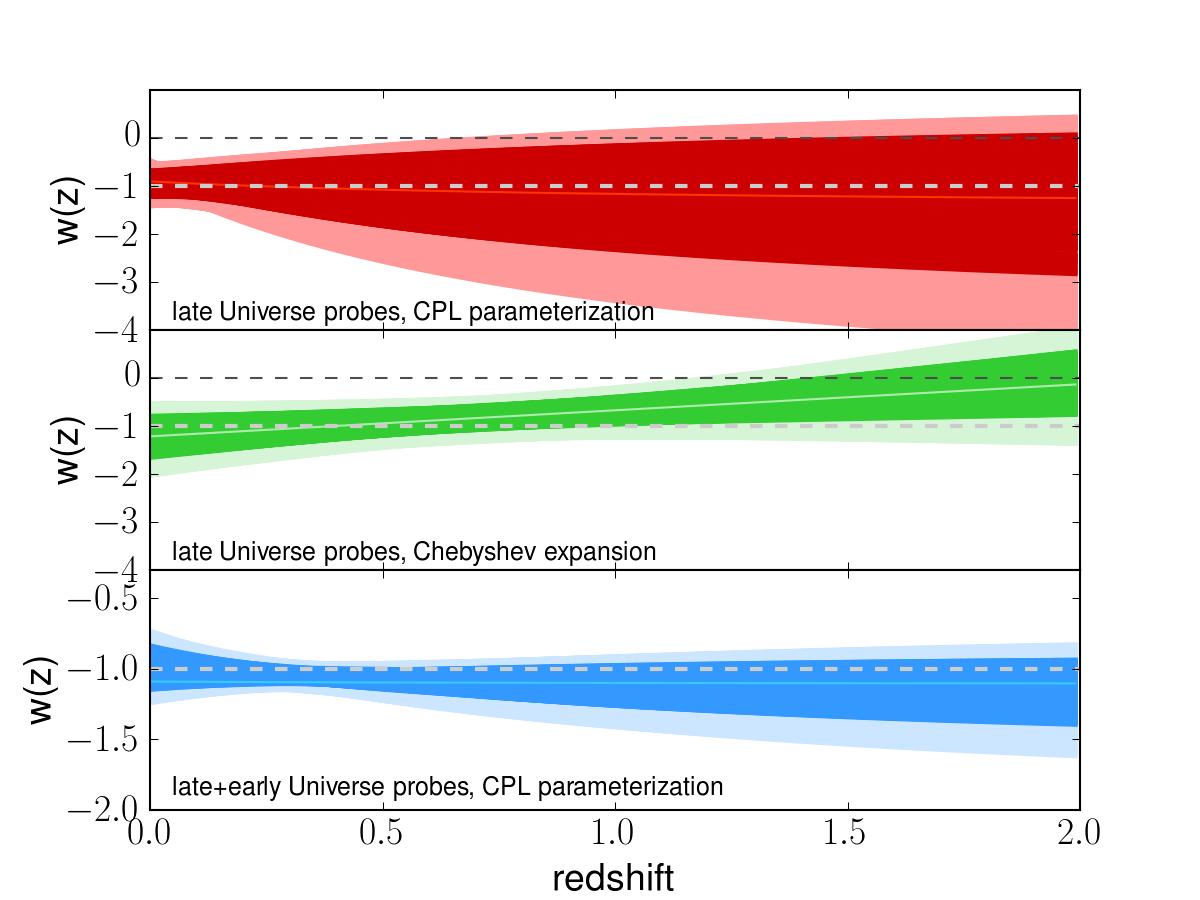}
\caption{Reconstruction of the time evolution of dark energy EoS obtained from different probes. Red contours the constraints obtained from 
late-Universe probes with a CPL parameterisation (see Eq. \ref{eq:Hztheor} and Tab. \ref{tab:fw0waCDM}), green contours show the constraints
obtained from late-Universe probes with Chebyshev decomposition up to the first order (see Eq. \ref{eq:Cheb1} and Tab. \ref{tab:Cheb}), and blue 
contours the constraints obtained adding also CMB information with the CPL parameterisation 
(see Eq. \ref{tab:fw0waCDM_Pl}). The darker regions show in each case the  68\% confidence level contours, while the light ones the 
 95\% confidence level contours. These constraints assume $H_{0}=73\pm2.4$ km/s/Mpc, taken from Ref.~\cite{Riess2011,Humphreys2013,Cuesta2015}.
\label{fig:wz}}
\end{center}
\end{figure}

CC measurements are extremely useful in breaking CMB parameters degeneracies for models beyond the ``minimal'' $\Lambda$CDM model.
In particular we will show their fundamental importance in models that for allow a time-variability of the dark energy equation of state and for models 
where CMB constraints show parameter degeneracies with the expansion history.

We start extending the ``minimal" $\Lambda$CDM where the additional parameter affects directly the late-time expansion history. 
We first explore a flat $w_0w_a$CDM model which constraints are presented in Fig.~\ref{fig:fig1} and Tab. 
\ref{tab:fw0waCDM_Pl}. For the flat $w_{0}w_{a}$CDM, the base analysis provided by the Planck collaboration include Planck15 and BAO data; as demonstrated in
\S \ref{sec:reslate}, CC and BAO have a similar constraining power on the $w_{0}-w_{a}$ plane, hence when combining Planck15+BAO+CC we obtain
a little gain in the constraints, still however being able to close the  95\% confidence level contours for $w_{a}$. 
When including also $H_0$, the errorbars shrink considerably, and we obtain $w_{0}=-1.09^{+0.19}_{-0.21}$ and $w_a = -0.02^{+0.61}_{-0.49}$ (at  68\% confidence level);
in comparison, the combination Planck15+BAO+CC+SNe+H$_{0}$ obtains $w_{0}=-0.98\pm 0.11$ and $w_{a}=-0.30^{+0.42}_{-0.34}$ (at  68\% confidence level). 

In Fig. \ref{fig:wz} we show the evolution of $w(z)$ up to $z=2$ (which is the maximum redshift covered by our datasets) for different parameterisations.
We compare the constraints obtained from Chebyshev reconstruction and from a CPL parameterisation from late-Universe probes alone, and from a CPL
parameterisation combining also Planck15 measurements. For homogeneity with CPL, we show here the results from Chebyshev reconstruction
up to the first order, to compare results with similar degrees of freedom. 
We find that all the best-fit constraints show deviations from $\Lambda$CDM only up to $\sim25\%$ up to $z=2$ for the CPL parameterization,
and up to $\sim60$\% up to $z=1.5$ for the Chebyshev reconstruction. Quite remarkably, all measurements are compatible at 2$\sigma$
with a constant dark energy EoS $w=-1$ in the entire redshift range. The most accurate measurement is obtained when combining both late- and
early-Universe probes together, providing a constraint on $w(z)$ which is consistent with the $\Lambda$CDM model at the 40\% level over the entire 
redshift range $0 < z < 2$ (at 68\% confidence level).

\begin{figure}[t!]
\begin{center}
\includegraphics[width=0.9\textwidth]{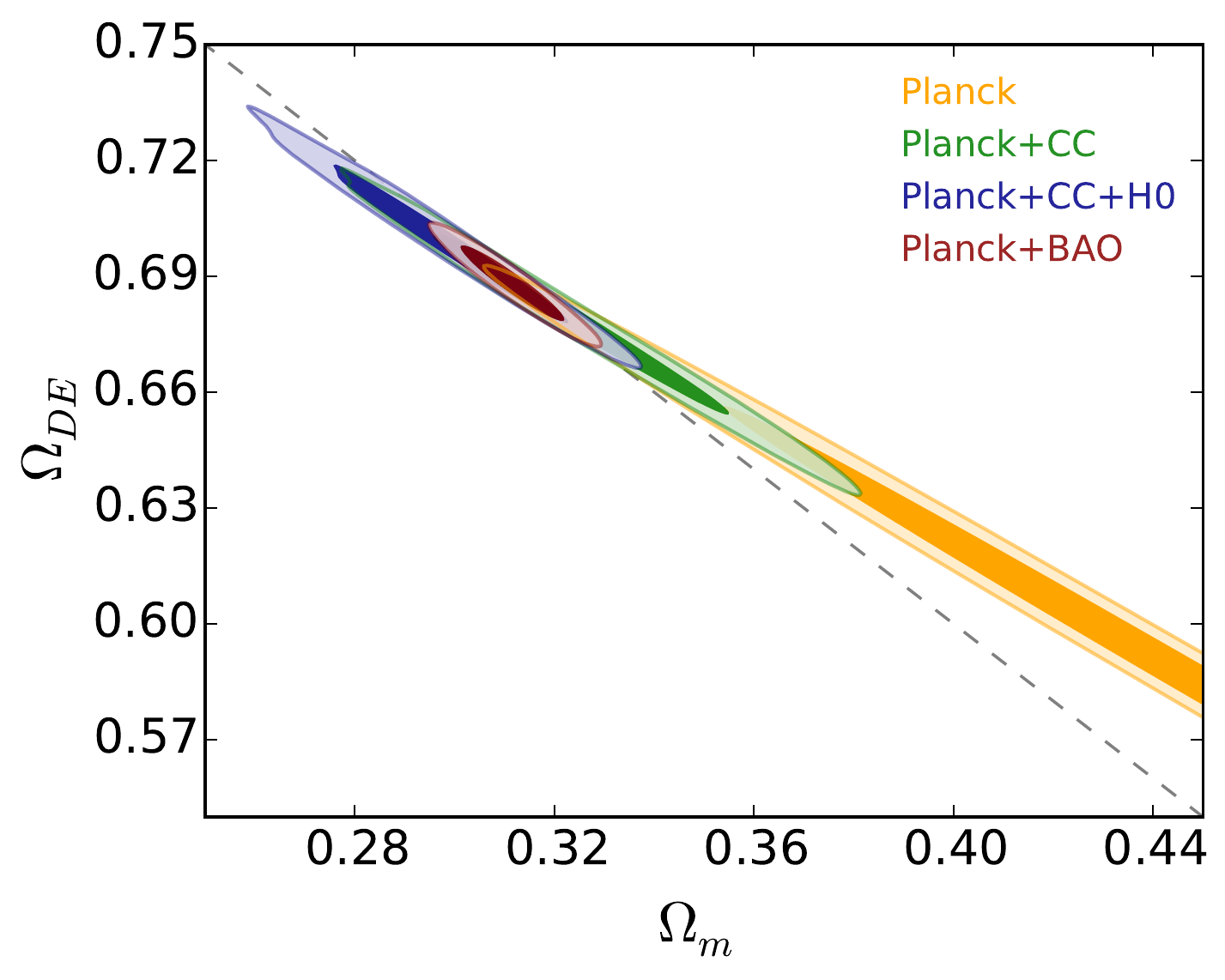}
\caption{Constraints in the $\Omega_{\rm DE}$-$\Omega_{\rm m}$ plane for different data-sets combinations. Note that combining
cosmic chronometer data-set with CMB data helps in constraining the contours close to the $\Omega_{\rm k}=0$ region, indicated with the dashed line. Similar results are obtained when combining the BAO with the CMB. We used $H_{0}=73\pm2.4$ km/s/Mpc from Ref.~\cite{Riess2011,Humphreys2013,Cuesta2015}.
\label{fig:fig2}}
\end{center}
\end{figure}

We then consider an open $\Lambda$CDM model, to explore the constraints that can be obtained on curvature from the different probes. The results
are shown in Fig.~\ref{fig:fig2} and Tab. \ref{tab:omk}. We find that cosmic chronometer are fundamental to break the $\Omega_{\rm m}$-$\Omega_{\rm DE}$
degeneracy present in CMB only data, constraining $\Omega_{k}=-0.0030^{+0.0055}_{-0.0042}$. The error bars can be reduced when also the $H_{0}$ 
data is included, obtaining $\Omega_{k}=0.0028^{+0.0035}_{-0.0031}$. For comparison, BAO data provide a similar result, $\Omega_{k}=0.0002\pm 0.0021$.

\begin{table}[b!]
\begin{center}
\begin{tabular}{cccc}
\multicolumn{4}{c}{\small MARGINALIZED 1D CONSTRAINTS}\\
\multicolumn{4}{c}{o$\Lambda$CDM model}\\
\hline \hline
& $\Omega_{\rm m}$ & $\Omega_{\rm DE}$ & $\Omega_{k}$\\
\hline
Planck15 & $0.47^{+0.063}_{-0.086}$ & $0.57^{+0.064}_{-0.047}$ & $-0.040^{+0.023}_{-0.016}$ \\
Planck15+CC & $ 0.325^{+0.018}_{-0.023}$ & $0.678^{+0.018}_{-0.015}$ & $-0.0030^{+0.0055}_{-0.0042}$ \\
Planck15+CC+H$_0$ & $ 0.298\pm 0.015$ & $ 0.699\pm 0.013$ & $0.0028^{+0.0035}_{-0.0031}$ \\
Planck15+BAO & $ 0.3117\pm 0.0070$ & $0.6881\pm 0.0065$ & $0.0002\pm 0.0021$ \\
\hline \hline
\end{tabular}
\end{center}
\caption{Constraints on $\Omega_{\rm m}$, $\Omega_{\rm DE}$ and $\Omega_{k}$ at  68\% confidence level obtained for an open $\Lambda$CDM cosmology. 
For $H_{0}$, we considered the value $73\pm2.4$ km/s/Mpc from Ref.~\cite{Riess2011,Humphreys2013,Cuesta2015}.}
\label{tab:omk}
\end{table}

\begin{figure}[t!]
\begin{center}
\includegraphics[width=0.9\textwidth]{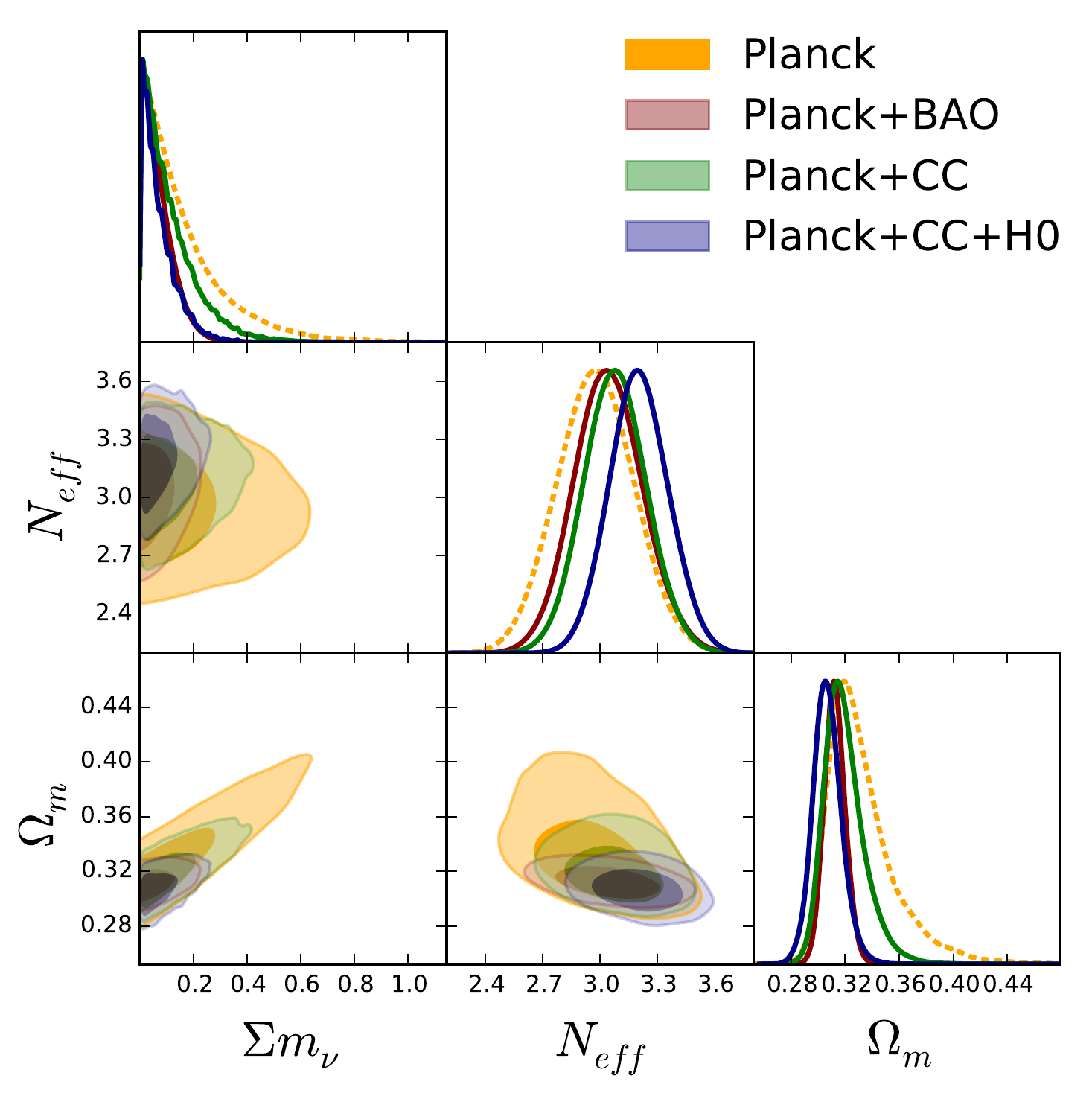}
\caption{Constraints on number of effective relativistic species $N_{\rm eff}$ and their total mass $\Sigma m_{\nu}$ obtained with different combination of probes. 
Note that cosmic chronometers data-set in combination with the CMB provide comparable constraints to those of CMB+BAO. We adopted $H_{0} = 73\pm2.4$ km/s/Mpc from Ref.~\cite{Riess2011,Humphreys2013,Cuesta2015}.
\label{fig:fig3}}
\end{center}
\end{figure}

\begin{table} [t!]
\begin{center}
\begin{tabular}{ccc}
\multicolumn{3}{c}{\small MARGINALIZED  (95\%) 1D CONSTRAINTS}\\
\multicolumn{3}{c}{$\Lambda$CDM+$\Sigma m_{\nu}$ model}\\
\hline \hline
& $\Sigma m_{\nu}$ & $N_{\rm eff}$\\
\hline
Planck15 & $< 0.577$ & $2.93^{+0.39}_{-0.38}$\\
Planck15+CC & $< 0.423$ & $3.05^{+0.33}_{-0.32}$\\
Planck15+CC+$H_0$ & $<0.269$ & $3.17^{+0.29}_{-0.30}$\\
Planck15+BAO & $< 0.222$ & $2.99^{+0.37}_{-0.34}$\\
\hline \hline
\end{tabular}
\end{center}
\caption{Constraints on $\Sigma m_{\nu}$ (in units of eV) and $N_{\rm eff}$ at  95\% confidence level obtained for a flat $\Lambda$CDM cosmology.
For $H_{0}$, we assumed a Gaussian prior $73\pm2.4$ km/s/Mpc from Ref.~\cite{Riess2011,Humphreys2013,Cuesta2015}.}
\label{tab:mnus8}
\end{table}

It is possible to constrain neutrino properties from the combination of CMB data with late-Universe measurement
of the expansion rate (e.g., see Refs.~\cite{Reid2010,DeBernardis2008,Riess2011}). In this work, we consider a simple flat $\Lambda$CDM 
Universe where the effective number of relativistic species $N_{\rm eff}$ is let free, and not fixed to 
the standard value of $N_{\rm eff}=3.04$; we analyze this model studying the combined dataset of Planck15 and CC and $H_0$.

Results are shown in Fig. \ref{fig:fig3}. Planck15+CC have the same constraining power as Planck15+BAO. They restrict $N_{\rm eff} = 3.17^{+0.29}_{-0.30}$. 
Clearly, CC are not able to measure $N_{\rm eff}$ directly, but provide constraints by breaking the degeneracy between the number of relativistic species 
and the parameters that fix the matter-radiation equality and the expansion history (e.g., see Ref. \cite{DeBernardis2008}).
We note how, in this case, almost all the statistical power of constraining $N_{\rm eff}$ is given by CC.
We then fit a flat $\Lambda$CDM Universe where the total neutrino mass is left as a free parameter, and looked at the constraints
on the sum of neutrino masses\footnote{We assume three degenerate neutrino species; current data have no sensitivity on the hierarchy, so this is a very 
good approximation.} We obtain a constraint of $M_{\nu} < 0.27$ at 95\% confidence.

\begin{figure}[t!]
\begin{center}
\includegraphics[width=0.9\textwidth]{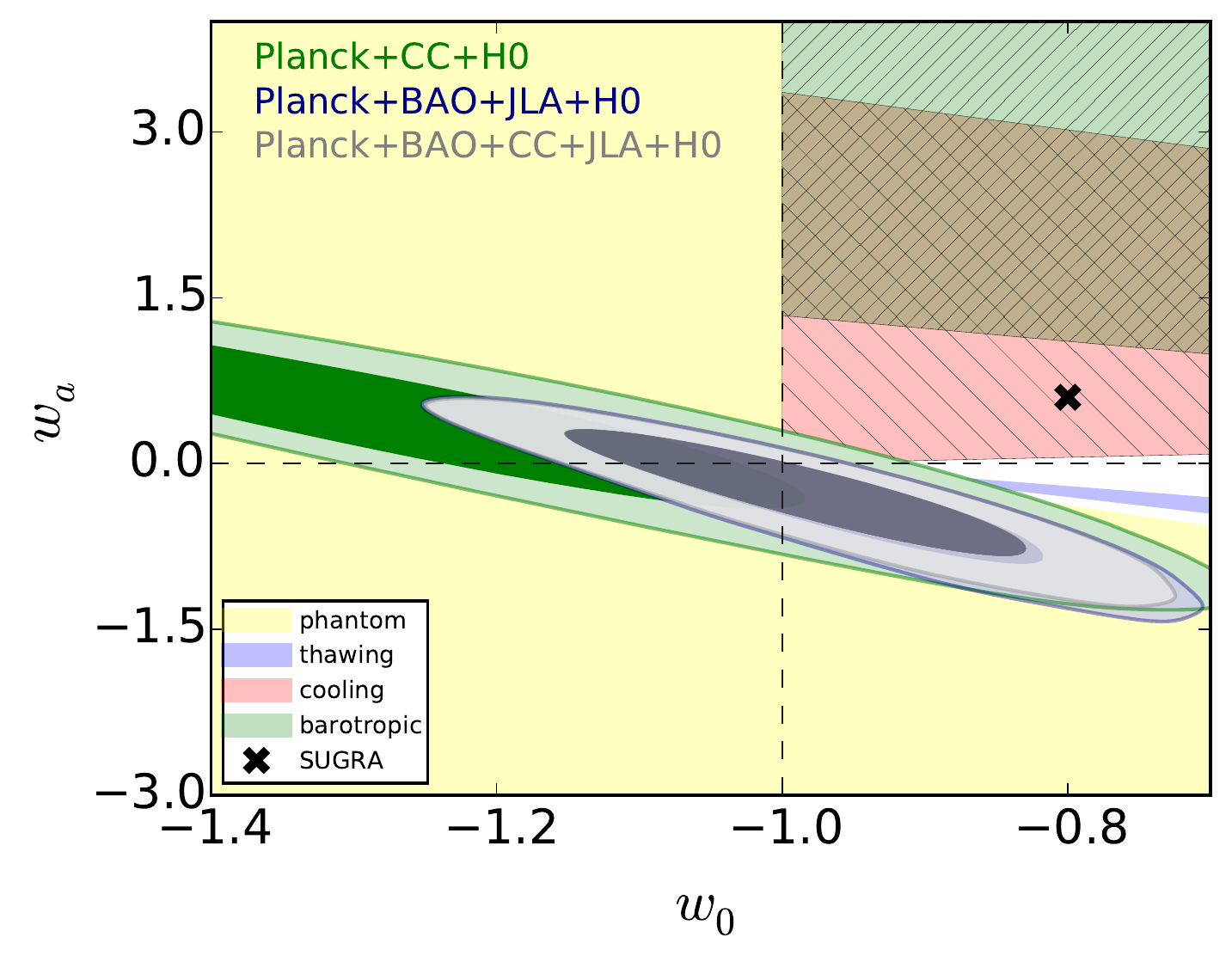}
\caption{Constraints on the time evolution of dark energy equation of state compared to several theoretical models (the brown region is the overlap between barotropic and cooling quintessence models). Models have been taken from
Refs.~\cite{Barger2006,Linder2003}. Dashed lines correspond to the position of the cosmological constant, $(w_{0},w_{a})=(-1,0)$.
The Hubble constant considered for these constraints is $H_{0}=73\pm2.4$ km/s/Mpc from Ref.~\cite{Riess2011,Humphreys2013,Cuesta2015}.
\label{fig:wz_models}}
\end{center}
\end{figure}

Finally, to exploit the constraining power of our new results in the $w_{0}-w_{a}$ plane, in Fig. \ref{fig:wz_models} we compare our measurements 
with some theoretical models taken from Ref.~\cite{Barger2006,Linder2003}. They comprise a large variety of dark energy models, including barotropic models,
phantom models, two different quintessence models and a SUGRA model. We refer to Refs.~\cite{Caldwell2005,Barger2006} for a comprehensive 
description of the models. Here we note that thank to the newly analysed data, the 95\% allowed region in the $w_{0}-w_{a}$ plane is significantly reduced. 
The new constraints allow not only to reject at 95\% confidence level the barotropic models, but also to exclude almost all quintessence models and the SUGRA
model.

\section{Conclusions}
\label{sec:conclusions}

Measurements of the expansion history are the only direct indication that the Universe is undergoing an accelerated expansion,  
therefore suggesting  that it should be dominated by a dark energy component \cite{Moresco2016}. 
Given the magnitude of the implications of the accelerated expansion, it is important to have 
different, independent and complementary observations of the expansion history. 

In this work we have used the latest cosmic chronometers measurements to set constraints on the time evolution of dark energy. 
We have shown that cosmic chronometers alone are able to set constraints on cosmological parameters
comparable to other state-of-the-art late-Universe probes, namely BAO and SNe. 
However, unlike BAO and SNe, the cosmic chronometer data provide a direct measurements of the expansion rate $H(z)$, 
without having to assume a cosmological model, and thus can be used to test it. 

Remarkably, we find that constraints on dark energy are fully consistent with a cosmological constant up to $z\sim2$, irrespectively on the assumed
parameterisation for $w(z)$, and when combining late- and early-Universe probes we find deviations from the $\Lambda$CDM model smaller than 40\% 
over the entire redshift range $0 < z < 2$ (at 68\% confidence level). In the CPL parameterisation, 
from the combination of late-Universe probes alone we find $w_0 = -0.9 \pm 0.18$ and $w_a = -0.5 \pm 1.7$, and when combining also CMB 
measurements from Planck15 we obtain $w_0=-0.98\pm 0.11$ and $w_a=-0.30\pm0.4$.
We compared these constraints with a set of theoretical dark energy models, finding that quintessence is disfavoured by the data
at high significance, and providing strong constraints on the allowed families of models.
We also tightly constrain the geometry of the Universe, obtaining a density parameter for curvature $\Omega_k = 0.003 \pm 0.003$.

Finally, we explored the power of cosmic chronometer in breaking degeneracies between parameters, and hence to constrain quantities that do not directly affects
the expansion history of the Universe, namely the number of relativistic species and the total mass of neutrinos. Our measurements provide a value
$N_{\rm eff} = 3.17 \pm 0.15$ that excludes at more than $5 \sigma$ the presence of an extra sterile neutrino, putting also a limit on the sum 
of neutrino masses, $\Sigma m_{\nu}< 0.27$ eV at 95\% confidence level.

Future surveys such as Euclid \cite{Laurejis2011}, WFIRST \cite{Spergel2013}, DESI \cite{DESI} and LSST \cite{LSST} will represent the next significant 
step, providing sub-percent measurements of the dark energy EoS, and giving fundamental insights for our comprehension of the elusive nature of dark energy.

\acknowledgments{MM and AC acknowledge contracts ASI-Uni Bologna-Astronomy Dept. Euclid-NIS I/039/10/0, 
and PRIN MIUR ``Dark energy and cosmology with large galaxy surveys''. LV and RJ acknowledge support by AYA2014-58747-P and MDM-2014- 0369 of 
ICCUB (Unidad de Excelencia Maria de Maeztu).}

\bibliographystyle{ieeetr}
\bibliography{bib}

\appendix
\section{Dependence of the results on the assumed evolutionary stellar population synthesis model}
\label{sec:modeldep}

\begin{figure}[t!]
\begin{center}
\includegraphics[width=0.49\textwidth]{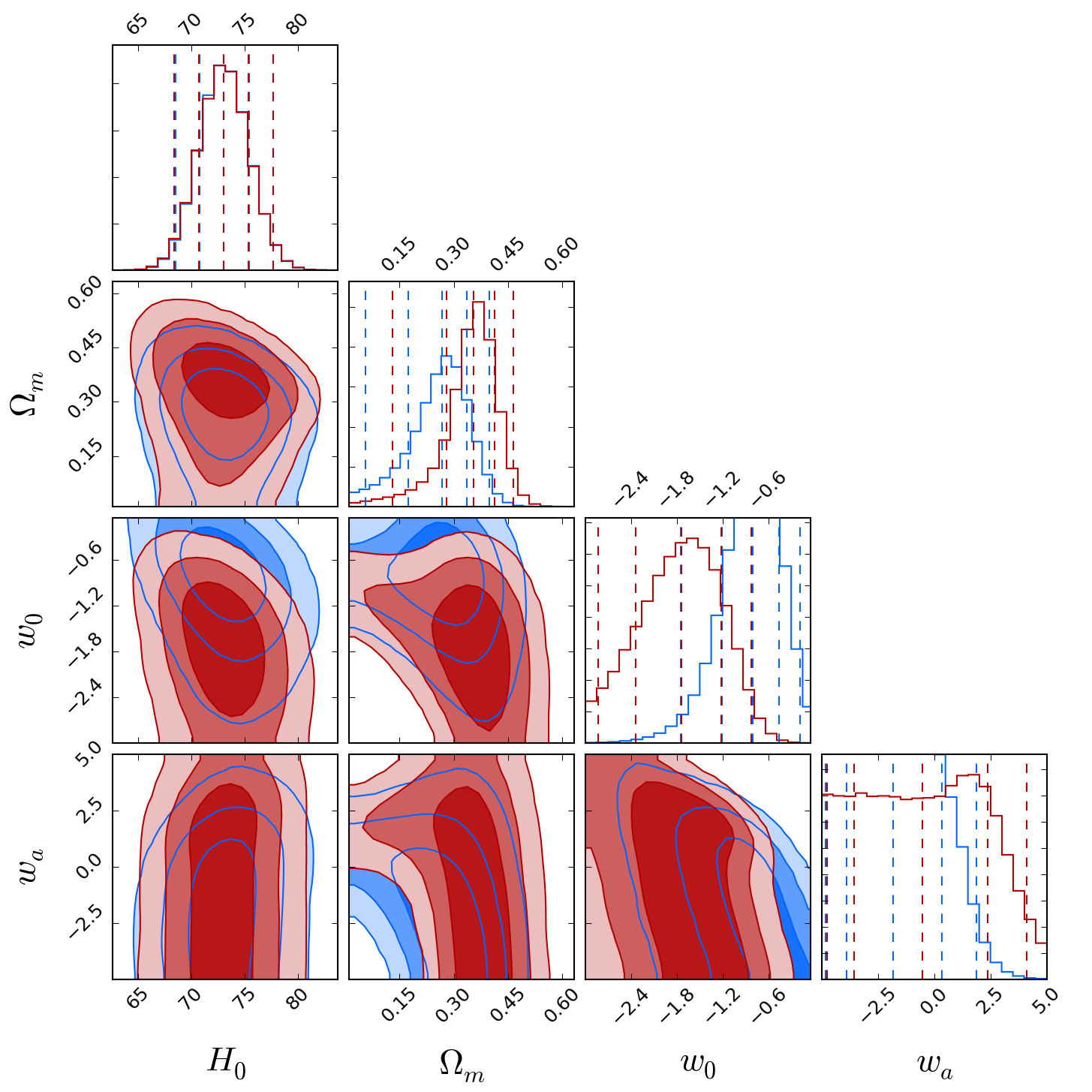}
\includegraphics[width=0.49\textwidth]{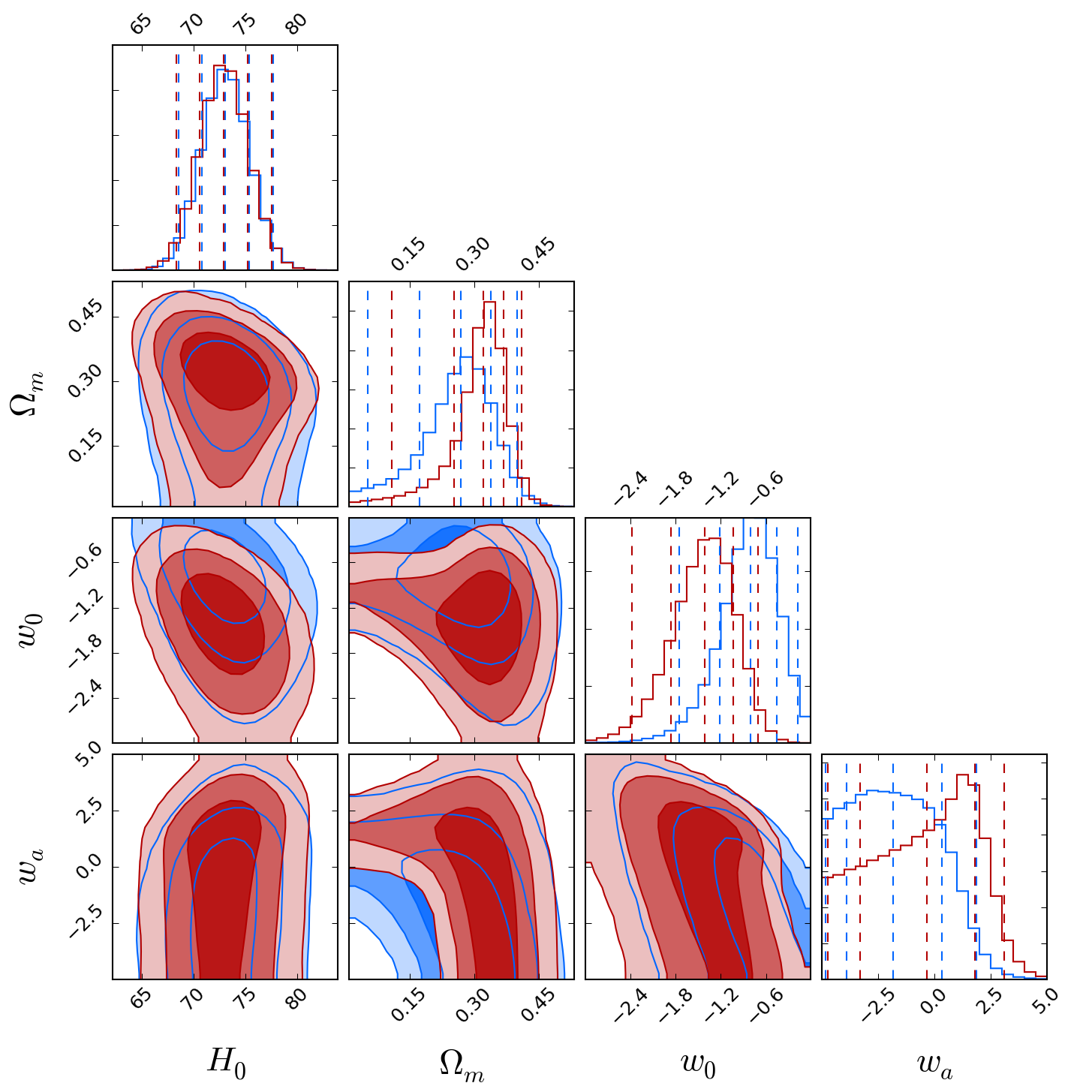}
\caption{Comparison of the results obtained fitting CC with BC03 (in red) and M11 models (in blue). In the left panel are shown
the constraints using measurements taken only from Refs.~\cite{Moresco2012a,Moresco2015,Moresco2016}, while in the right
panel are also considered additional measurements obtained with BC03 model from Refs.~\cite{Simon2005,Stern2010,Zhang2014}.
For this constraints, we assumed a Gaussian prior on $H_{0}=73\pm2.4$ km/s/Mpc from Ref.~\cite{Riess2011,Humphreys2013,Cuesta2015}.
\label{fig:BCvsM11}}
\end{center}
\end{figure}

\begin{table}[h!]
\begin{center}
\begin{tabular}{ccccc}
\multicolumn{5}{c}{\small MARGINALIZED 1D CONSTRAINTS}\\
\multicolumn{5}{c}{f$w_{0}w_{a}$CDM model}\\
\hline \hline
& H$_{0}$ & $\Omega_{\rm m}$ & $w_{0}$ & $w_{a}$\\
\hline
BC03 \cite{Moresco2012a,Moresco2015,Moresco2016} & $74.2^{+20.1}_{-15.8}$ & $0.33^{+0.26}_{-0.19}$ & $-1.86^{+1.61}_{-1.1}$ & $-0.8^{+5.2}_{-4}$\\
M11 \cite{Moresco2012a,Moresco2015,Moresco2016} & $78.3^{+19.6}_{-15.2}$ & $0.23^{+0.18}_{-0.18}$ & $-1.19^{+1.1}_{-1.5}$ & $-1.1^{+4.7}_{-3.7}$\\
BC03 \cite{Simon2005,Stern2010,Moresco2012a,Zhang2014,Moresco2015,Moresco2016} & $73.4^{+19.2}_{-13.4}$ & $0.3^{+0.18}_{-0.2}$ & $-1.46^{+1.27}_{-1.4}$ & $-0.5^{+4.5}_{-4.2}$\\
\hline
BC03 \cite{Moresco2012a,Moresco2015,Moresco2016}+H$_{0}$ & -- & $0.36^{+0.11}_{-0.23}$ & $-1.73^{+0.9}_{-1.1}$ & $-0.5^{+4.6}_{-4.2}$\\
M11 \cite{Moresco2012a,Moresco2015,Moresco2016}+H$_{0}$ & -- & $0.27^{+0.13}_{-0.21}$ & $-0.81^{+0.6}_{-0.9}$ & $-1.8^{+3.7}_{-3}$\\
BC03 \cite{Simon2005,Stern2010,Moresco2012a,Zhang2014,Moresco2015,Moresco2016}+H$_{0}$ & -- & $0.32^{+0.09}_{-0.21}$ & $-1.4^{+0.7}_{-1}$ & $-0.3^{+3.4}_{-4.4}$\\
\hline
\end{tabular}
\end{center}
\caption{Comparison of the constraints obtained by fitting CC assuming different EPS (BC03 and M11) 
on H$_{0}$, $\Omega_{\rm m}$, $w_{0}$ and $w_{a}$ at  95\% confidence level obtained for a flat $\Lambda$CDM 
cosmology with a CPL parameterisation for the dark energy EoS. In the lower part of the table, we assumed a Gaussian 
prior on $H_{0}=73\pm2.4$ km/s/Mpc from 
Ref.~\cite{Riess2011,Humphreys2013,Cuesta2015}, and hence do not provide the constraints for $H_{0}$.}
\label{tab:BC03vsM11}
\end{table}

To measure the Hubble parameter $H(z)$ within the CC approach, an evolutionary stellar population synthesis (EPS) model has to be assumed, to calibrate the relation
between age and age-related observables used in our analysis. Amongst the models available in literature, so far BC03 \cite{BC03}
and M11 \cite{MaStro} have been used, encompassing significant differences in stellar phases considered, methods implemented and
models adopted; for a more detailed discussion, we refer to Ref.~\cite{Moresco2016}. In the following, we explore the dependence of
our results on the assumed EPS model.

As a reference, we consider the flat $w_{0}w_{a}$CDM model, and fitted CC data obtained assuming BC03 and M11 models. 
We performed two different test: firstly, we consider only the limited dataset for which both BC03 and M11 measurements are available
\cite{Moresco2012a,Moresco2015,Moresco2016}, and then we compare the result obtained for the full BC03 sample, i.e by considering also
the measurements from Refs. \cite{Simon2005,Stern2010,Zhang2014} (for these measurements, only BC03 constraints are available).
We also explored how the constraints change by adding the local measurements of $H_{0}$, as a pivot value for the $H(z)$ relation.

The results are shown in Fig. \ref{fig:BCvsM11}, and reported in Tab. \ref{tab:BC03vsM11}.
We find no significant difference between the two constraints, with only M11 measurements pointing to a value of
$H_{0}$ at higher odds than BC03 with respect to state-of-art measurement. The measurements are still well compatible
also when a Gaussian prior on $H_{0}$ is assumed, with the full BC03 dataset providing tighter constraints on cosmological
parameters, as expected given the higher number of data available. 

We therefore conclude that our measurements are robust, and do not significantly depend on the assumed EPS model.

\end{document}